\documentclass[%
  reprint,
  superscriptaddress,
  amsmath,amssymb,
  aps,prl,
 ]{revtex4-1}
\usepackage{subfigure}
\usepackage{graphicx} 
\usepackage{todonotes}

\newcommand{\avg}[1]{\ensuremath{\left<#1\right>}}

\begin{document}

\author{Gabriele Tocci}
\email{gabriele.tocci@epfl.ch}
\affiliation{Laboratory for fundamental BioPhotonics, Institutes of Bioengineering and Materials Science and Engineering, School of Engineering, and Lausanne Centre for Ultrafast Science,  {\'E}cole Polytechnique F{\'e}d{\'e}rale de Lausanne (EPFL), CH-1015 Lausanne, Switzerland}
\affiliation{Laboratory of Computational Science and Modeling, Institute of Materials, {\'E}cole Polytechnique F{\'e}d{\'e}rale de Lausanne, 1015 Lausanne, Switzerland}
\author{Chungwen Liang}
\affiliation{Laboratory for fundamental BioPhotonics, Institutes of Bioengineering and Materials Science and Engineering, School of Engineering, and Lausanne Centre for Ultrafast Science,  {\'E}cole Polytechnique F{\'e}d{\'e}rale de Lausanne (EPFL), CH-1015 Lausanne, Switzerland}
\affiliation{Laboratory of Computational Science and Modeling, Institute of Materials, {\'E}cole Polytechnique F{\'e}d{\'e}rale de Lausanne, 1015 Lausanne, Switzerland}
\author{David M. Wilkins}
\affiliation{Laboratory for fundamental BioPhotonics, Institutes of Bioengineering and Materials Science and Engineering, School of Engineering, and Lausanne Centre for Ultrafast Science,  {\'E}cole Polytechnique F{\'e}d{\'e}rale de Lausanne (EPFL), CH-1015 Lausanne, Switzerland}
\affiliation{Laboratory of Computational Science and Modeling, Institute of Materials, {\'E}cole Polytechnique F{\'e}d{\'e}rale de Lausanne, 1015 Lausanne, Switzerland}
\author{Sylvie Roke}
\affiliation{Laboratory for fundamental BioPhotonics, Institutes of Bioengineering and Materials Science and Engineering, School of Engineering, and Lausanne Centre for Ultrafast Science,  {\'E}cole Polytechnique F{\'e}d{\'e}rale de Lausanne (EPFL), CH-1015 Lausanne, Switzerland}
\author{Michele Ceriotti}
\affiliation{Laboratory of Computational Science and Modeling, Institute of Materials, {\'E}cole Polytechnique F{\'e}d{\'e}rale de Lausanne, 1015 Lausanne, Switzerland}%

\title{Second Harmonic Scattering as a \\Probe of Structural
Correlations in Liquids}



\begin{abstract}
Second harmonic scattering
experiments of water and other bulk
molecular liquids
have long been assumed to be insensitive to
interactions between the
molecules. The measured
intensity is generally thought to arise
from incoherent scattering due to individual
molecules. We introduce a method to compute the
second harmonic scattering pattern of
molecular liquids directly from atomistic
computer simulations, which 
takes into account the coherent
 terms. 
We apply this approach to
large scale molecular dynamics
simulations of liquid water,
where we show that nanosecond second harmonic
scattering  experiments contain a coherent
contribution arising from
radial and angular correlations
on a length-scale of $\lesssim 1$~nm - much
shorter than had been recently hypothesized~\cite{sheltonJCP2014}.
By combining structural 
correlations from simulations with
experimental data~\cite{sheltonJCP2014}
we can also extract an effective
molecular hyperpolarizability in the
liquid phase. This work demonstrates 
that second harmonic
scattering experiments and atomistic simulations
can be used in synergy to investigate the
structure of complex liquids, solutions and biomembranes, 
including the intrinsic inter-molecular 
correlations.
\end{abstract}

\maketitle
Nonlinear light scattering has been widely used to
investigate aqueous interfaces, including
suspensions of metallic or semiconducting nanoparticles,
water droplets and biological membranes
(for example, Refs.~\cite{eisenthalchemrev2006,rokegonellarev2012,gonella2012effect,
schurer2010probing,butet2010optical,russier2010symmetry,singh2013second,gonella2016second,smolentsev2016intermolecular,
chen2015three,yan1998new,yang2001angle,russier2004wavelength,kauranen1995supramolecular}).
In particular, second harmonic scattering (SHS), also commonly referred to
as Hyper-Rayleigh scattering (HRS; the terms SHS and HRS are used interchangeably
in this work), is especially sensitive to the molecular orientation
of liquids at interfaces: there is a greater contribution to
radiated second harmonic light arising from molecules 
at the interface, where inversion symmetry is broken,
compared to those in the bulk, which is on average centrosymmetric.
In bulk liquids
the coherent term is usually assumed to be
negligible and the measured signal is
 mostly interpreted in terms of 
fully incoherent scattering
from individual
molecules~\cite{terhunePRL1965,kauranen1996theory}.
Under this assumption, the SHS intensity of 
a solution can be expressed as
a linear combination of the incoherent 
contribution of the
solute and of that of the solvent.
Using this approach the average
hyperpolarizability of a molecule in 
solution can be estimated~\cite{ClaysPRL1991}.

This assumption stems from early work by Terhune
\textit{et al.}, who neglected the
coherent contribution to the intensity because
the wavelength of
the laser is much larger than the range of
molecular correlations expected in a liquid~\cite{terhunePRL1965}.
Although 
Bersohn and Maker later
developed a theory to include
coherent scattering due to
structural correlations~\cite{bersohn1966double, maker1970spectral}, these 
additional terms have been 
largely ignored in the interpretation
of experiments. 
More recently, in a series of articles
by Shelton \textit{et al.},
a sign of coherence has been
reported from nanosecond HRS measurements
on bulk liquids including water and acetonitrile~\cite{shelton2000polarization,shelton2002collective,shelton2005dipolar,sheltonJCP2014,shelton2015long}.
A diverse range of possible origins was suggested
for this coherence: a third order response
arising from collective
polar modes~\cite{shelton2000polarization},
collective rotation of
molecules~\cite{shelton2002collective},
the presence of ferroelectric
domains~\cite{shelton2005dipolar}, coupling
of rotations and translations in acoustic phonons extending up to 2000 nm~\cite{sheltonJCP2014}, 
or orientational correlations
found in ideal isotropic homogeneous random vector
fields~\cite{shelton2015long}.
%
%
Clearly SHS measurements contain information
about the intermolecular structure of liquids
although it is evident that no consensus has been
reached on their origin (see \textit{e.g.} Ref.~\cite{pounds2007dipolar}) and
their length scale.
So far, experiments in bulk water have been
interpreted by treating intermolecular correlations
as fitting parameters~\cite{sheltonJCP2014,shelton2015long}
and a framework to explicitly account
for the correlations
that give rise to coherent scattering in SHS
experiments has not yet been developed.
Recent elastic femtosecond SHS
measurements from
ionic solutions have shown that several different electrolytes
induce long-range orientational correlations in 
water, starting at concentrations as low
as 10 $\mu$M~\cite{chen2016electrolytes}.
The strong dependence on isotopic composition
also suggests a link between these observations and 
the H-bond network of water.

%

The liquid-phase hyperpolarizability plays a
crucial role in the
interpretation of SHS experiments.
Measurements have often been analyzed using a
molecular non-linear response tensor of water
obtained from quantum chemical
calculations~\cite{gubskaya2001multipole}.
However, the model that underlies the evaluation of
of the hyperpolarizability in Ref.~\cite{gubskaya2001multipole}
described the surroundings of a water molecule by
three point charges, which does not take into account the many
complex molecular environments that can be found in liquid water~\cite{Gasparotto2016}.
%
%
%

In this work we present a method
to compute the SHS intensity directly from 
atomistic computer simulations,
including both the
incoherent  \textit{and} the coherent
contributions. 
This framework is used to calculate the
angular scattering pattern from large-scale
force-field molecular dynamics (MD) simulations of
liquid water.
We show that quantitative agreement with nanosecond
HRS experiments~\cite{sheltonJCP2014} cannot be
achieved by using
the values of molecular hyperpolarizability
often used to interpret experiments,~\cite{gubskaya2001multipole}
obtained with M{\o}ller-Plesset
perturbation theory expanded up
to the fourth order (MP4).
We find instead that it is possible to
obtain satisfactory agreement by combining the
correlations obtained from simulations
with an effective liquid phase
molecular hyperpolarizability,  
used as a fitting parameter. 
The values obtained for this hyperpolarizability
can be used to provide a more 
quantitative interpretation of other SHS
experiments. By computing explicitly
the correlations that contribute to the 
SHS signal, we can also gather insight
into the length scales that are most
relevant for these experiments. 

We begin by introducing the 
general expression for
the SHS intensity $I(\mathbf{q},\mathbf{u},
\mathbf{v})$
for an ensemble of $N$ molecules
at a scattered wave-vector
$\mathbf{q}$ and with the
polarization directions of the incoming
and outgoing beams defined by the vectors $\mathbf{u}$ and $\mathbf{v}$:
\begin{align}
\label{eq:tot_int}
I(\mathbf{q},\mathbf{u},
\mathbf{v})\propto N^{-1}
\left \langle \left
|\sum_{\alpha}  
\beta(\alpha,
\mathbf{u},\mathbf{v})
e^{\mathrm{i} 
\mathbf{q}\cdot\mathbf{r}_{\alpha}
}\right |^2 \right \rangle ,
\end{align}
where the brackets indicate a time average over
uncorrelated configurations and
$\beta(\alpha, \mathbf{u},
\mathbf{v})$ is the component of
the hyperpolarizability 
of a scattering unit $\alpha$
in the laboratory reference
frame (L) projected 
onto the polarization
direction of the incoming
($\mathbf{u}$) and outcoming
($\mathbf{v}$) beams,
\textit{i.e.} $ \beta(\alpha, \mathbf{u},
\mathbf{v}) = \sum_{LMN}
\beta^{\textrm{L}}_{LMN}(\alpha)v_L u_M
u_N$, and the proportionality constant can be found
in Ref.~\cite{bersohn1966double}.
A schematic of the geometric setup is shown
in the Supporting Information (SI) in Fig.~S1.
Eq.~\ref{eq:tot_int} can be decomposed
into the sum of an incoherent term 
and of a coherent term 
respectively~\cite{bersohn1966double}:
\begin{equation} 
\label{eq:int_sum}
\begin{split}
&I(\mathbf{q},\mathbf{u},
\mathbf{v}) 
\propto N^{-1}
\left \langle  \sum_{\alpha}
\beta(\alpha,
\mathbf{u},\mathbf{v})^2
\right \rangle + \\
&N^{-1}
\left \langle \sum_{\alpha}  
\sum_{\alpha^{\prime} \neq \alpha }
\beta(\alpha,
\mathbf{u},\mathbf{v})
\beta(\alpha^{\prime},
\mathbf{u},\mathbf{v})
e^{[\mathrm{i} 
\mathbf{q}\cdot(\mathbf{r}_{\alpha} - 
\mathbf{r}_{\alpha^{\prime}})
]} \right \rangle .
\end{split}
\end{equation}
The hyperpolarizability in the
laboratory frame $\beta^{\textrm{L}}_{LMN}$
can be obtained
by applying the rotations
 $\beta^{\textrm{L}}_{LMN}=\sum_{abc} \beta^{\textrm{M}}_{abc} c_{aL} c_{bM}
c_{cN}$, with
$\beta^{\textrm{M}}_{abc}$ being
the molecular hyperpolarizability
in the molecular frame and
$c_{iI}$ the projection of the
$i^{\rm th}$ unit vector in the
 molecular frame on the $I^{\rm th}$ unit vector in the lab frame.
 In the following we assume
 that $\beta^{\textrm{M}}_{abc}$ can be expressed
 as an effective molecular hyperpolarizability
 $\avg{\beta^{\textrm{M}}_{abc}}$.

As previously discussed, in bulk liquids
the coherent term in Eq.~\ref{eq:int_sum} 
-- which describes the interference between the
waves scattered by two molecules -- 
would average to zero if the instantaneous relative 
orientation of different molecules were completely
random, and is often assumed to be negligible compared to the incoherent
term~\cite{eisenthalchemrev2006,terhunePRL1965,maker1970spectral,ClaysPRL1991}
In order to test this assumption,
we developed a framework to explicitly evaluate
Eq.~\ref{eq:tot_int} for
MD simulations.
We applied this framework to MD simulations
of liquid water involving
about 260,000 
molecules, using a cubic simulation
box with a side of 20 nm.
Simulations were performed with the GROMACS code v.5.0.4 \cite{gromacsv5},
using the rigid TIP4P/2005 water model \cite{tip4p2005}.
The equations of motion were integrated in the NVT
ensemble using the velocity Verlet algorithm for 20 ns
with a 2 fs timestep.
Temperature control was achieved using the stochastic velocity
rescaling thermostat~\cite{bussi_don}, with a target temperature of 300 K.
Full details of the computational
setup can be found in the Supporting Information.
We performed extensive tests
on the setup to ensure that it was
insensitive to finite size effects and the type of force-field used.

\begin{figure}
{\includegraphics[width=0.42\textwidth]{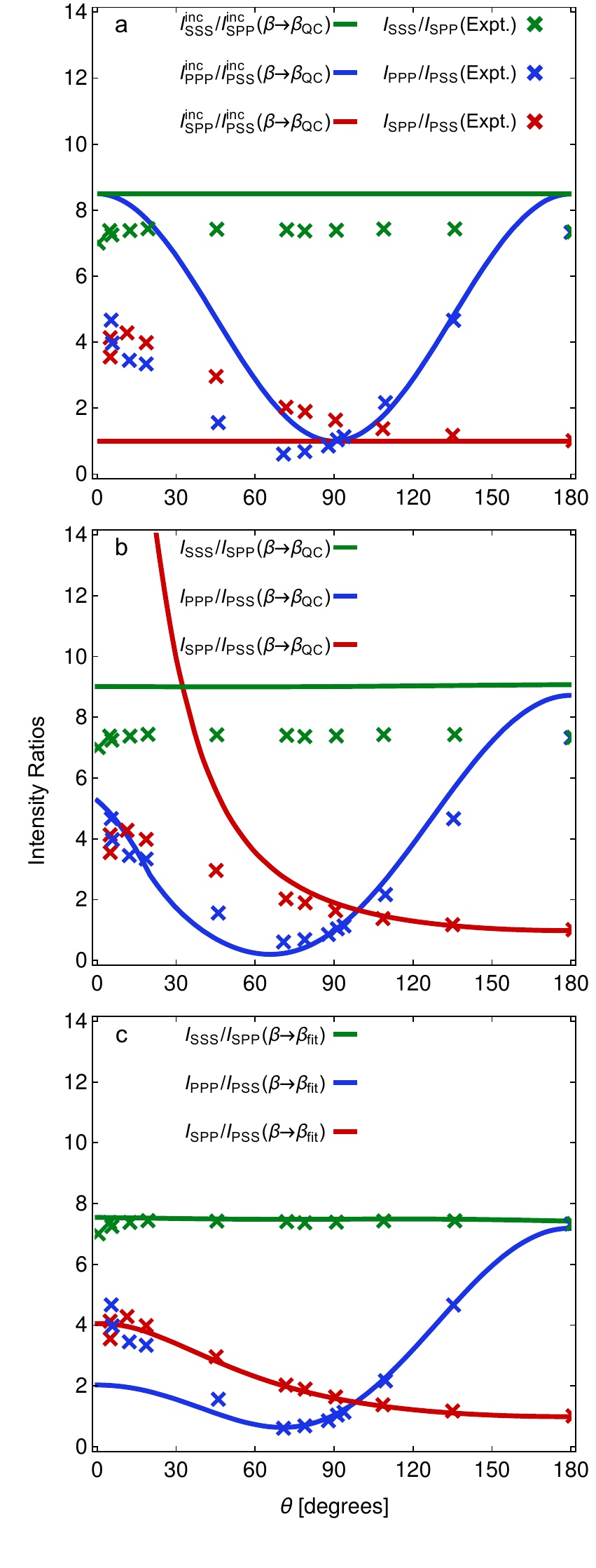}
\caption{SHS intensity ratios
of bulk water for different polarization combinations as
a function of the scattering angle
$\theta$ computed from MD (solid lines)
and as measured by Shelton in Ref.~\cite{sheltonJCP2014}
(crosses).
Intensity ratios
with the  molecular
hyperpolarizability of H$_2$O in
the liquid from Ref.~\cite{gubskaya2001multipole}
(\textit{i.e.} 
$\beta \rightarrow \beta_{\rm QC}$)
and (a) assuming incoherent scattering,
(b) including also the coherent contribution.
(c) Ratios obtained from
Eq.~\ref{eq:int_fitbeta} including 
both the incoherent and the coherent
contributions and where the
molecular hyperpolarizability has been
fitted to experiment,
\textit{i.e.} $\beta \rightarrow \beta_{\rm fit}$.
The nanosecond HRS experimental data is reproduced with
permission from the American
Institute of Physics from Ref.~\cite{sheltonJCP2014}.
}
\label{fig:intensities}}
\end{figure}

Fig.~\ref{fig:intensities}
shows the ratios of the SHS intensities
of bulk water reported by Shelton in
Ref.~\cite{sheltonJCP2014} using nanosecond experiments at
different polarization combinations as a
function of the scattering angle 
($\theta$) and as computed in this work from
the analysis of our MD simulations.
The convention used for the polarization
combinations 
follows Ref.~\cite{chen2016electrolytes}
and references therein. Light polarized parallel to the scattering plane is
denoted by P and light polarized perpendicular to the scattering plane is denoted
by S. A polarization combination is
specified as XYY, with X (= P or S) the polarization of the outgoing
beam and Y (= P or S) that of the incoming beam.
Fig.~\ref{fig:intensities}(a)
illustrates the intensity ratios 
obtained from our simulations
under the assumption of purely
incoherent scattering. We use the constant
hyperpolarizability from Ref.~\cite{gubskaya2001multipole},
computed using MP4
and modelling a liquid-like environment
by placing point charges in a
geometry resembling the first
solvation shell.
Despite its simplicity,
this model has been used in
several studies of the 
hyperpolarizability of water \cite{ward2013second,sonoda2005simulation}.
It can be seen that
a model for the intensity
assuming only incoherent
scattering 
completely fails to account for the qualitative
features of the experimental curves, which show an asymmetry
with respect to 90 degrees and a varying $I_\text{SPP}/I_\text{PSS}$ ratio.

We then proceed to 
compute the intensity with the full
Eq.~\ref{eq:tot_int}, using the 
$\avg{\beta^\text{M}_{abc}}$ elements from
Ref.~\cite{gubskaya2001multipole}.
As shown in
Fig.~\ref{fig:intensities}(b),
the functional form of
the plot extracted from the
simulation is markedly different
from that of 
Fig.~\ref{fig:intensities}(a)
and reproduces the main
features of the experimental curves.
However, the computed ratios show a clear
quantitative difference from the experimental
ratios.
Rather than trying to refine the evaluation of 
this effective hyperpolarizability~\cite{sonoda2005simulation,sylvester2004sign,kongstedJCP2004}, one
could extract the molecular correlations  
from the atomistic simulation, and verify whether 
the experimentally measured intensities
can be reproduced by using the
values of $\avg{\beta^\text{M}_{abc}}$  as fitting parameters. 

In order to investigate this point,
we reformulate Eq.~\ref{eq:tot_int} in a form
that separates intermolecular correlations and
the tensor elements of the constant effective
molecular hyperpolarizability $\langle
\beta^{\textrm{M}}_{abc} \rangle$ in the liquid phase.
We introduce the sixth-order tensor $ W_{abcdef} (\mathbf{q},\mathbf{u},
\mathbf{v}) 
= 
N^{-1} \sum_{\alpha, \alpha^{\prime}} S^{*}_{abc}(\mathbf{q},\mathbf{r}_{
\alpha},\mathbf{u},
\mathbf{v})
S_{def}(\mathbf{q},\mathbf{r}_{
\alpha^{\prime}},\mathbf{u},
\mathbf{v}) $, where,
\begin{equation}
\label{eq:Stens}
S_{abc}(\mathbf{q},\mathbf{r}_{
\alpha},\mathbf{u},
\mathbf{v})=  \sum_{LMN} c_{aL} v_L c_{bM} u_M c_{cN} u_N
\exp{(\mathrm{i} 
\mathbf{q}\cdot\mathbf{r}_{\alpha}
)}.
\end{equation}
\noindent $ W_{abcdef} (\mathbf{q},\mathbf{u},
\mathbf{v})$ includes both the
single-molecule term giving rise to
incoherent scattering and
the coherent contribution due to
the intermolecular radial and
angular correlations. Inserting this expression
into Eq.~\ref{eq:tot_int} we obtain,
\begin{equation}
\label{eq:int_fitbeta}
I(\mathbf{q},\mathbf{u},
\mathbf{v}) \propto \sum_{abcdef} \left \langle \beta^{\textrm{M}}_{abc} \right \rangle \left \langle \beta^{\textrm{M}}_{def} \right \rangle
\left \langle 
W_{abcdef} (\mathbf{q},\mathbf{u},
\mathbf{v}) \right
\rangle.
\end{equation}
%
Because SHS experiments are performed at a value of
$q$ of the order of $10^{-2}$ nm$^{-1}$, which is
too small to be probed in simulations,
we extrapolate $\avg{W_{abcdef} (\mathbf{q},\mathbf{u},
\mathbf{v})}$ to the $q\rightarrow 0~\text{nm}^{-1}$
limit (see SI).
Having separated the expression for the intensity
into the effective molecular
hyperpolarizability and the structural
correlation term  $\langle W_{abcdef} (\mathbf{q},\mathbf{u},\mathbf{v})\rangle$, it is possible
to use Eq.~\ref{eq:int_fitbeta} 
to determine the value of $\langle 
\beta^{\textrm{M}}_{abc} \rangle $ 
that best matches experiments.
The $\avg{\mathbf{W} (\mathbf{q},\mathbf{u},
\mathbf{v})}$ tensor needs to be computed only once, and 
weighted with tentative values of the 
hyperpolarizability tensor to
find the best fit.
Under nonresonant conditions~\cite{giordmainePR1965},
and under the assumption that each water 
molecule in the condensed phase has $C_{2v}$ symmetry,
there are only three independent
components of the molecular
hyperpolarizability tensor,
\textit{i.e.} $\beta^{\textrm{M}}_{zxx}$,
$\beta^{\textrm{M}}_{zyy}$ and $\beta^{\textrm{M}}_{zzz}$.
One value of the three elements of the
hyperpolarizability is linearly
dependent on the other two
because most of the time, intensity ratios are extracted from SHS experiments,
rather than the bare intensities.

\begin{figure}
{\includegraphics[width=0.4\textwidth]{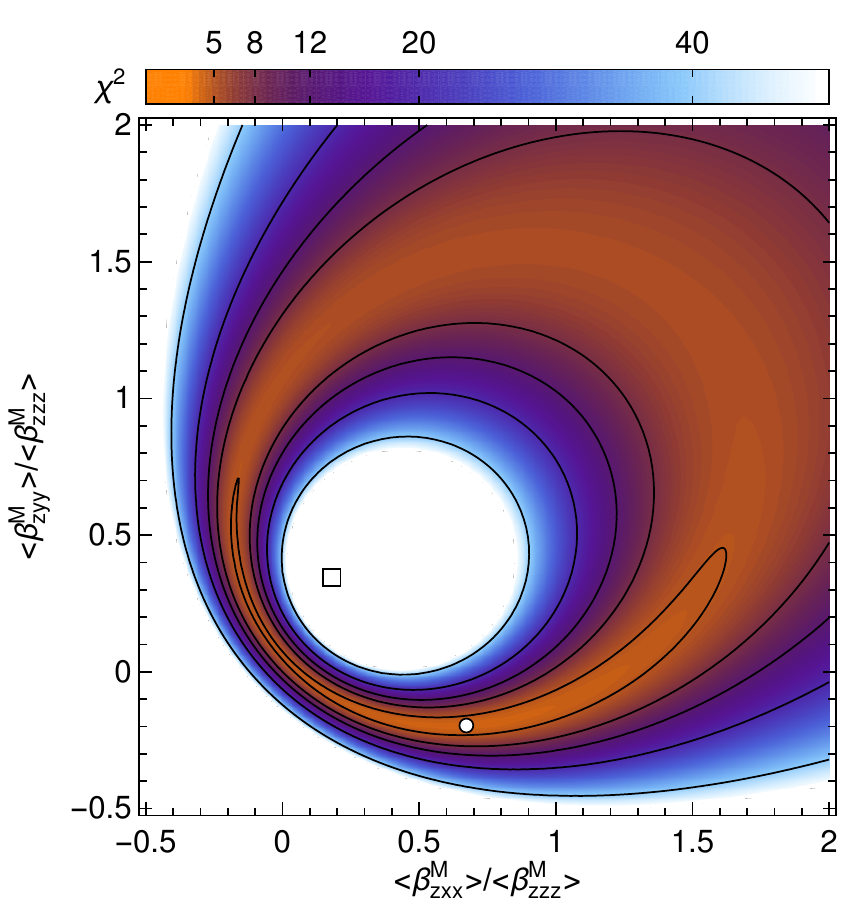}
\caption{Contour plot of the mean
standard error of the simulated
intensity ratios from the experimental
data ($\chi^2$)
as a function of the 
independent components of the
effective hyperpolarizability
of water in the liquid phase.
The circle and the square symbol
indicate the values of the 
hyperpolarizabilities
that give the minimum in the error
and those computed in
Ref.~\cite{gubskaya2001multipole},
respectively.}
\label{fig:contours_beta_error}}
\end{figure}

With these hypotheses in place, Fig.~\ref{fig:contours_beta_error}
shows the error ($\chi^2$) of the fit 
to the experimental data of Ref.~\cite{sheltonJCP2014}
as a function of
$\langle \beta^{\textrm{M}}_{zxx} \rangle/\langle 
\beta^{\textrm{M}}_{zzz} \rangle$ and 
$\langle \beta^{\textrm{M}}_{zyy} \rangle/\langle 
\beta^{\textrm{M}}_{zzz} \rangle$. 
This plot can be used to ascertain the quality
of a given model to compute
the effective molecular
hyperpolarizability
in the condensed phase.
For instance, the values of the molecular
hyperpolarizabilities in
the model of
Ref.~\cite{gubskaya2001multipole} (the square
symbol)
give a much larger $\chi^2$ compared
to the best estimate obtained by
our fitting procedure
(the circle symbol).
It can also be seen that
the parameter space of $\langle \beta^{\textrm{M}}_{zxx} \rangle/\langle 
\beta^{\textrm{M}}_{zzz} \rangle$ and 
$\langle \beta^{\textrm{M}}_{zyy} \rangle/\langle 
\beta^{\textrm{M}}_{zzz} \rangle$ that results in a small error
is rather broad (see the orange region with $\chi^2 < 8$ in
Fig.~\ref{fig:contours_beta_error}).
Combining the data of Ref.~\cite{sheltonJCP2014}
with that of other experiments,
\textit{e.g.} on electrolyte 
solutions~\cite{chen2016electrolytes},
might allow one to narrow down the uncertainty
on $\langle 
\beta^{\textrm{M}}_{abc} \rangle $.
Fig.~\ref{fig:intensities}(c)
shows directly the agreement
between the measured intensity ratio and
the results obtained from
Eq.~\ref{eq:int_fitbeta}
using our best estimate for 
$\langle \beta^{\textrm{M}}_{abc} \rangle $,
combined with the
structural intermolecular correlations
obtained from the MD trajectories.
The reason for the discrepancy between the experimental and the 
simulated intensity ratios in Fig.~\ref{fig:intensities}(b) 
can thus be ascribed to the different value used for the 
effective hyperpolarizability.
Although Fig.~\ref{fig:intensities}(c)
agrees well with experiments,
the agreement is not perfect at low scattering
angles. The underlying reason
is most likely the fact that the scattering
plane becomes ill-defined as $\theta\rightarrow 0$, because the two vectors 
defining this plane ($\mathbf{k}_{\rm in}$ and $\mathbf{k}_{\rm out}$, 
as defined in the SI) become collinear in this limit.




%
%

\begin{figure}
{\includegraphics[width=0.4\textwidth]{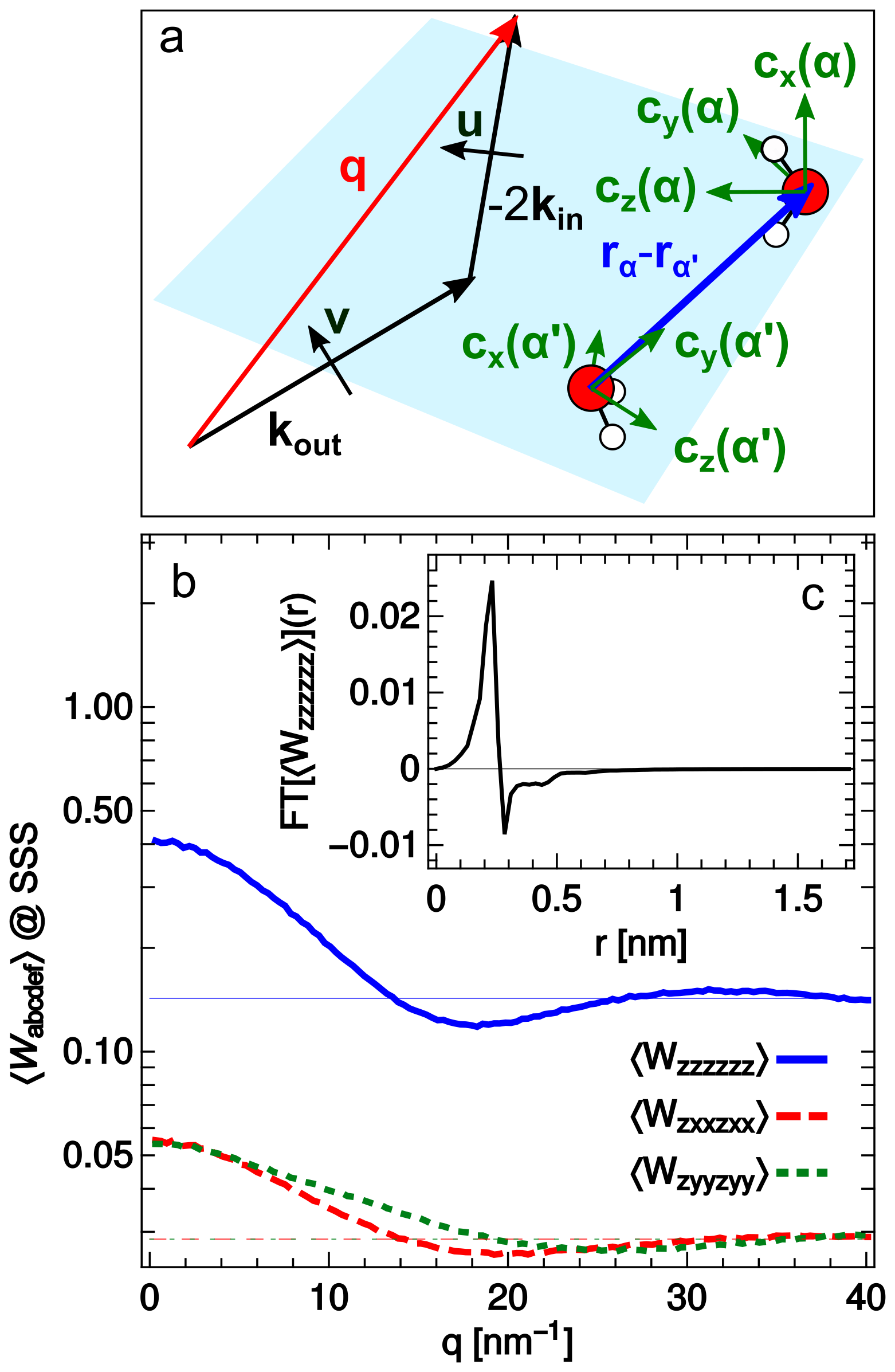}
\caption{The role of angular and radial correlations in determining the 
SHS intensity. (a) Schematic of the SHS system in an arbitrary geometry 
including for simplicity only a pair
of molecules, where $\mathbf{q}$,
$\mathbf{k}_{\mathrm{in}}$, $\mathbf{k}_\mathrm{{out}}$
$\mathbf{u}$ and $\mathbf{v}$ define the experimental setup, while the
$c$ vectors are the unit vectors of molecule $\alpha$ and $\alpha^{\prime}$
in their respective reference frames. (b) Wave vector dependence of 
$\langle W_{zzzzzz} \rangle$, $\langle W_{zxxzxx} \rangle$ and $\langle W_{zyyzyy} \rangle$ tensor elements in the SSS polarization (thin lines are the incoherent parts of these tensors). (c) Fourier transform of $\langle W_{zzzzzz} \rangle(q)$ demonstrating the short-range nature of the correlations that contribute to SHS.}
\label{fig:W_q_SSS}}
\end{figure}

Fig.~\ref{fig:W_q_SSS} provides
further insight into the correlations probed by SHS experiments.
Fig.~\ref{fig:W_q_SSS}(a) shows a schematic of
the experimental system, including for simplicity only a pair
of molecules. SHS probes the correlations
between projections of unit vectors in the molecular frame onto those
of the laboratory reference frame for a given orientation of $\mathbf{q}$
and combination of $\mathbf{u}$ and $\mathbf{v}$. Fig.~\ref{fig:W_q_SSS}(a) thus
provides a pictorial representation of Eqs.\ref{eq:Stens} and \ref{eq:int_fitbeta}.
Different components of the $\avg{\mathbf{W}(\mathbf{q,\mathbf{u},\mathbf{v}})}$ tensor report on complex
combinations of radial and angular correlations. As a concrete example, 
$\langle W_{zzzzzz} \rangle$ for the SSS polarization 
combination, taking an experimental setup in which $\mathbf{q}$ 
lies in the $XY$ plane, is given explicitly by
$\langle\sum_{\alpha,\alpha^{\prime}} c_{zZ}^{3}(\alpha) 
c_{zZ}^{3}(\alpha^{\prime}) e^{i \mathbf{q}\cdot\left(\mathbf{r}_{\alpha} - 
\mathbf{r}_{\alpha^{\prime}}\right)}\rangle$,
which involves the pair correlation function of the cube of 
the molecular dipole's component in the direction of the  
laboratory $Z$ axis.
Fig.~\ref{fig:W_q_SSS}(b) shows the effect of these 
correlations on three components of
$\avg{\mathbf{W}(\mathbf{q,\mathbf{u},\mathbf{v}})}$ as a function of wavevector. In the limit as $q\rightarrow 0$
there is a significant coherent contribution, which is a hallmark of
intermolecular correlations.\cite{bersohn1966double}
It can be seen that in this 
limit $\langle W_{zzzzzz} \rangle$ 
takes the largest value, 
while $\langle W_{zxxzxx} \rangle$ and $\langle W_{zyyzyy} \rangle$ are 
about 8 times smaller. 
This reflects the fact that the 
correlations between the 
water dipoles are stronger than those 
between the $x$ and $y$ 
molecular axes.

In the limit as $q\rightarrow 0$, $\avg{\mathbf{W}(\mathbf{q,\mathbf{u},\mathbf{v}})}$ reports on the magnitude of 
the correlations integrated over all
distances. By computing
the value of $\avg{\mathbf{W}(\mathbf{q,\mathbf{u},\mathbf{v}})}$ at finite $q$
we can quantify the length scale
that gives the largest
contribution to the SHS signal. 
As shown in Figure~\ref{fig:W_q_SSS}(b), 
$\langle W_{zzzzzz} \rangle$ converges to the 
incoherent limit, oscillating 
with a period of about 20 nm$^{-1}$, which 
corresponds to a strong contribution from 
correlations
within the first and second solvation 
shell of each water molecule.
This is shown in more detail by the Fourier transform of 
$\avg{W_{zzzzzz}(q)}$ into the spatial domain, given in
Fig.~\ref{fig:W_q_SSS}(c).
The pair correlation function shown in the figure presents
two pronounced peaks, corresponding to the first two
solvation shells, and a decays to zero within
$\sim 1 $~nm.
Therefore, the largest contribution to  $\avg{W_{zzzzzz}(q\rightarrow 0)}$ arises from pairs of
molecules within the first couple of solvation
shells, while long-distance
tails give a negligible contribution.
Correlations beyond a 
few nm are not crucial, as is clear from Fig.~S2, where
it is shown that the intensity ratios 
differ by a few percent when 
using a 5~nm simulation box compared to a 20~nm box.
The SHS patterns of Shelton~\cite{sheltonJCP2014}
 have been given a number of different interpretations,
all of which involve the role of long-range correlations. 
Although intermolecular
correlations on longer scales 
may be present, as observed in MD simulations of neat
water~\cite{zhang2014dipolar} and in SHS experiments (as 
well as MD simulations) of electrolyte 
solutions~\cite{chen2016electrolytes},
SHS measurements in neat 
water can be rationalized in terms of short-range 
correlations.

%
Fig.~\ref{fig:W_q_SSS}(b)
focuses on an experimental setup that
is particularly simple to interpret. However,  $\avg{\mathbf{W}(\mathbf{q,\mathbf{u},\mathbf{v}})}$
captures \emph{all} of the correlations that are
relevant for an SHS experiment. In the SI (see Figs.~S4
and~S5) we show the dependence 
of various tensor elements on the scattering angle
and for different polarization combinations. Depending
on the geometry of the experiment
and the values of $\avg{\mathbf{\beta}_{abc}^\text{M}}$, different measurements probe 
different combinations of molecular correlations. 
Based on this observation, it is possible
to design experiments that 
are particularly sensitive to a given kind of correlation. 
For instance, with PPP polarization,
$\theta\approx 0$ and 
$\theta\approx 180^\circ$ give the strongest coherent contribution, which is at odds with 
previous suggestions that measurements
should be performed at
$\theta = 90^\circ$.~\cite{terhunePRL1965}

Because SHS probes angular
intermolecular
correlations in liquids,
it can be used to gain further
insight into the structure
of non-centrosymmetric
molecular fluids in addition
to the more widely used
X-ray and neutron scattering,
which are mostly sensitive 
to radial intermolecular
correlations~\cite{clark2010small,sedlmeier2011spatial,amann2016x}.
While there have been extensive
computational investigations of X-ray scattering experiments,
the simulation of SHS experiments has received rather limited attention~\cite{janssen1999molecular}.
Previous experiments have been
analyzed under the assumption of
an incoherent SHS
intensity~\cite{yan1998new,russier2004wavelength,terhunePRL1965,maker1970spectral,ClaysPRL1991,ward2013second}.
For instance, under this
assumption the hyperpolarizability
of several solute molecules
in the condensed phase has been
extracted~\cite{ClaysPRL1991}.
Based on the developments presented here,
it is now also possible to analyze the 
signal from the solvent, without 
assuming that the scattering process is
incoherent, and being able to evaluate the SHS intensity for
arbitrary scattering angle, polarization combination and wave vector.
Furthermore, separating the molecular hyperpolarizability
$\langle \beta^{\textrm{M}}_{abc} \rangle$
and the structural term $\left \langle W_{abcdef} \right \rangle$
in \ref{eq:int_fitbeta} makes it possible to disentangle the role of
the second order optical response in the liquid
from that of the intermolecular correlations,
thereby allowing for a more thorough connection to experiments, and for
an unbiased assessment of the most 
relevant length scales.
An important question
for future work would be,
for instance, to investigate 
ionic solutions in order to evaluate
the nature and length scales of the 
intermolecular water-water correlations
that underlie the results of 
Ref.~\cite{chen2016electrolytes}.

In conclusion, we have developed a 
method that can be used in atomistic 
computer simulations of liquids
to calculate the second harmonic
scattering intensity, and which
fully accounts for the coherent
contribution to the scattering due to
interactions between molecules.
The method also provides a way
to extract an effective molecular
hyperpolarizability in the liquid
phase, without having to rely on
an over-simplified representation of the 
complex local
environment of a water
molecule~\cite{gubskaya2001multipole,kongsted2003nonlinear,sylvester2004sign}.
By applying this method to
nanosecond HRS experiments of liquid water~\cite{sheltonJCP2014}
we have shed light onto their coherent nature.
We have also provided deep insights into 
the radial and angular correlations
probed by these experiments, showing
that the angular dependence of the 
SHS signal can be explained in terms
of inter-molecular correlations on
a length scale of the order of 1 nm. 
The main assumption made in this work is that nanosecond SHS experiments
can be described in terms of 
an effective molecular
hyperpolarizability
$\avg{\beta^{\textrm{M}}_{abc}}$.
Work is currently underway to go beyond 
this treatment, by characterizing the 
dependence of $\beta^{\textrm{M}}_{abc}$
on the molecular environment, and 
thereby assessing the role of 
fluctuations.
Overall, we hope that these developments
will stimulate the use of
molecular simulations to aid the
interpretation of
SHS experiments
in more complex bulk liquids and at
aqueous interfaces.

\begin{acknowledgements}
GT, CL
and SR are grateful for support from the
Julia Jacobi Foundation, the Swiss National
Science Foundation (grant number
200021\_140472), and the European Research
Council (grant number
616305). DMW and MC acknowledge
funding from the Swiss National Science
Foundation
(Project ID 200021-159896). We are also
grateful for the generous allocation of CPU
time by CSCS under Project
ID s619.
\end{acknowledgements}

\section{Supporting Information}
    The Supporting Information contains
    further details of the following
    aspects: the geometrical setup of
    the SHS experiments; the computational details of the MD simulations; a discussion on the convergence of
	the SHS simulations at low scattering wavevectors; a discussion of the effect
	of finite-size systems and of different water force-field
	on the computation of the intensity ratios.
%
\bibliography{biblio,largebibliography} 
\end{document}